\documentclass[conference]{IEEEtran}
\usepackage[font=small,skip=6pt]{caption}
\usepackage{tikz}
\usetikzlibrary{calc,trees,positioning,arrows,chains,shapes.geometric, decorations.pathreplacing,decorations.pathmorphing,shapes, matrix,shapes.symbols}

\tikzset{
>=stealth',
  punktchain/.style={
    rectangle, 
    rounded corners, 
    draw=black, very thick,
    text width=10em, 
    minimum height=2.5em, 
    text centered, 
    on chain},
  punktchain2/.style={
    rectangle, 
    rounded corners, 
    draw=none, very thick,
    text width=10em, 
    minimum height=2.5em, 
    text centered, 
    on chain},
  line/.style={draw, thick, <-},
  element/.style={
    tape,
    top color=white,
    bottom color=blue!50!black!60!,
    minimum width=8em,
    draw=blue!40!black!90, very thick,
    text width=10em, 
    minimum height=3.5em, 
    text centered, 
    on chain},
  every join/.style={->, thick,shorten >=1pt},
  decoration={brace},
  tuborg/.style={decorate},
  tubnode/.style={midway, right=2pt},
}

\ifCLASSINFOpdf
\else
\fi
\begin{document}

%
\title{Technical Aspects of Cyber Kill Chain}

\author{\IEEEauthorblockN{Tarun Yadav}
\IEEEauthorblockA{Scientist, Defence Research and\\Development Organisation, INDIA\\Email: tarunyadav@hqr.drdo.in}

\and
\IEEEauthorblockN{Rao Arvind Mallari}
\IEEEauthorblockA{Scientist, Defence Research and\\Development Organisation, INDIA\\Email:arvindrao@hqr.drdo.in}}

\maketitle

\begin{abstract}

Recent trends in targeted cyber-attacks has increased the interest of research in the field of cyber security. Such attacks have massive disruptive effects on organizations, enterprises and governments. Cyber kill chain is a model to describe cyber-attacks so as to develop incident response and analysis capabilities. Cyber kill chain in simple terms is an attack chain, the path that an intruder takes to penetrate information systems over time to execute an attack on the target. This paper broadly categories the methodologies, techniques and tools involved in cyber-attacks. This paper intends to help a cyber security researcher to realize the options available to an attacker at every stage of a cyber-attack.
\end{abstract}

\IEEEpeerreviewmaketitle

\section{Introduction}
One of the leading problems faced by organizations is the emergence of targeted attacks conducted by adversaries who have easy access to sophisticated tools and technologies with an aim at establishing a persistent and undetected presence in the targeted cyber infrastructure. These multi-staged attacks are now becoming more complex, involving vertical and horizontal movement across multiple elements of the organization. The security research community has given this multi-stage chain of events culminating to cyber espionage a name: \textit{ The Cyber Kill Chain}. This paper aims on providing a primer on the cyber kill chain and surveys the recent trends and methodologies of the attacker at each stage of the cyber kill chain.

The paper talks about attacks in general irrespective of operating systems or application software since Cyber Kill Chain is a process rather than a technology. Technologies involved at each stage of  the cyber kill chain process  is explained without going into much details. 

The paper is organized into 4  sections. Section 2 introduces the phases of cyber kill chain, section 3 discusses the technical trends at each step of the cyber kill chain. The papers ends with concluding remarks in section 4.

\section{Cyber Kill Chain}
Cyber kill chain is a model for incident response team,digital forensic investigators and malware analysts to work in a chained manner. Inherently understanding Cyber kill chain is modeling and analyzing offensive actions of a cyber-attacker\cite{Weaponize:2}. So for a security analyst who develops defensive counter measures\cite{cyberkillchain:4}, it is of utmost importance to study the cyber kill chain. This knowledge can help one think on the same lines of that of an attacker. Each phase of kill chain in itself is a vast research area to tackle and analyze.

In recent years, cyber attacks have been more complex than it used to be and hence more destructive and dangerous\cite{Weaponize:7}\cite{ar10}. Nowadays multiple redundant attack vectors are being exploited in cyber attacks to not only multiply the effect but also making it more difficult for the response team to analyze. 

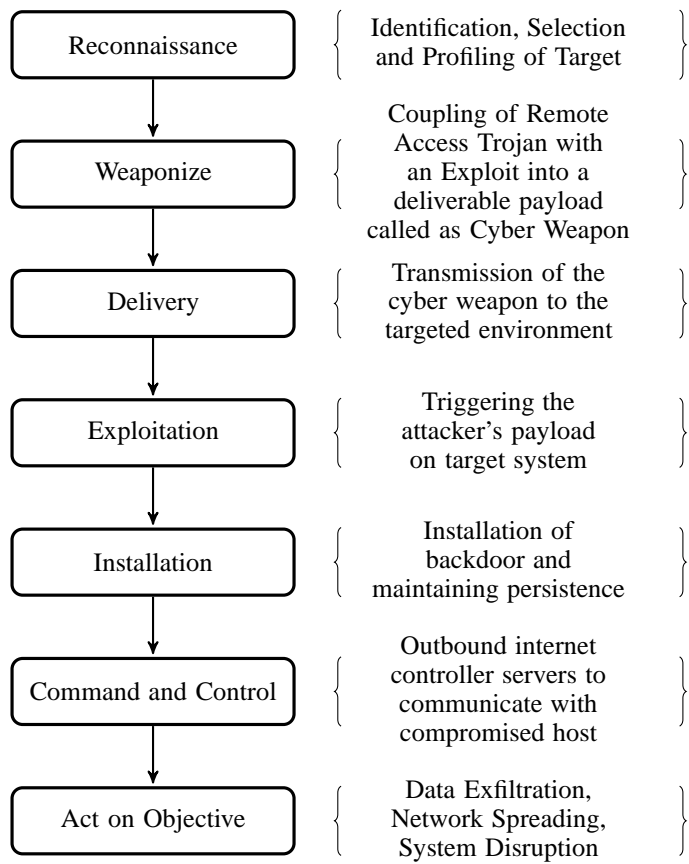
\begin{figure}

\begin{tikzpicture}
  [node distance=.8cm,
  start chain=going below,]
     \node[punktchain, join] (reconn) {Reconnaissance};
 \draw[tuborg, decoration={brace,mirror},xshift=.5cm] let \p1=(reconn.north), \p2=(reconn.south) in
    ($(2, \y1)$) -- ($(2, \y2)$) ;
\begin{scope}[start branch=branch1,]
      \node (des1) [punktchain2, on chain=going right] { Identification,  Selection and Profiling of Target};
    \end{scope}
\draw[tuborg, decoration={brace},xshift=5cm] let \p1=(reconn.north), \p2=(reconn.south) in
    ($(2, \y1)$) -- ($(2, \y2)$) ;
     \node[punktchain, join] (weapon)      {Weaponize};
\draw[tuborg, decoration={brace,mirror},xshift=.5cm] let \p1=(weapon.north), \p2=(weapon.south) in
    ($(2, \y1)$) -- ($(2, \y2)$) ;
\begin{scope}[start branch=branch2,]
      \node (des2) [punktchain2, on chain=going right] {Coupling  of Remote Access Trojan with an Exploit into a deliverable payload called as Cyber Weapon};
    \end{scope}
\draw[tuborg, decoration={brace},xshift=5cm] let \p1=(weapon.north), \p2=(weapon.south) in
    ($(2, \y1)$) -- ($(2, \y2)$) ;
     \node[punktchain, join] (del)      {Delivery};
\draw[tuborg, decoration={brace,mirror},xshift=.5cm] let \p1=(del.north), \p2=(del.south) in
    ($(2, \y1)$) -- ($(2, \y2)$) ;
\begin{scope}[start branch=branch3,]
      \node (des3) [punktchain2, on chain=going right] {Transmission of the cyber weapon to the targeted environment};
    \end{scope}
\draw[tuborg, decoration={brace},xshift=5cm] let \p1=(del.north), \p2=(del.south) in
    ($(2, \y1)$) -- ($(2, \y2)$);
\node[punktchain, join] (exploit) {Exploitation};
\draw[tuborg, decoration={brace,mirror},xshift=.5cm] let \p1=(exploit.north), \p2=(exploit.south) in
    ($(2, \y1)$) -- ($(2, \y2)$) ;
\begin{scope}[start branch=branch4,]
      \node (des4) [punktchain2, on chain=going right] {Triggering the attacker's payload on target system};
    \end{scope}
\draw[tuborg, decoration={brace},xshift=5cm] let \p1=(exploit.north), \p2=(exploit.south) in
    ($(2, \y1)$) -- ($(2, \y2)$);
     \node[punktchain, join] (install) {Installation};
\draw[tuborg, decoration={brace,mirror},xshift=.5cm] let \p1=(install.north), \p2=(install.south) in
    ($(2, \y1)$) -- ($(2, \y2)$) ;
\begin{scope}[start branch=branch5,]
      \node (des5) [punktchain2, on chain=going right] { Installation of backdoor and maintaining persistence};
    \end{scope}
\draw[tuborg, decoration={brace},xshift=5cm] let \p1=(install.north), \p2=(install.south) in
    ($(2, \y1)$) -- ($(2, \y2)$);
     \node[punktchain, join, ] (cnc) {Command and Control};
\draw[tuborg, decoration={brace,mirror},xshift=.5cm] let \p1=(cnc.north), \p2=(cnc.south) in
    ($(2, \y1)$) -- ($(2, \y2)$) ;
\begin{scope}[start branch=branch6,]
      \node (des6) [punktchain2, on chain=going right] {Outbound internet controller servers to communicate with compromised host};
    \end{scope}
\draw[tuborg, decoration={brace},xshift=5cm] let \p1=(cnc.north), \p2=(cnc.south) in
    ($(2, \y1)$) -- ($(2, \y2)$);
      \node [punktchain,join ](objectives)   {Act on Objective};
\draw[tuborg, decoration={brace,mirror},xshift=.5cm] let \p1=(objectives.north), \p2=(objectives.south) in
    ($(2, \y1)$) -- ($(2, \y2)$) ;
\begin{scope}[start branch=branch7,]
      \node (des7) [punktchain2, on chain=going right] { Data Exfiltration, Network Spreading, System Disruption};
    \end{scope}
\draw[tuborg, decoration={brace},xshift=5cm] let \p1=(objectives.north), \p2=(objectives.south) in
    ($(2, \y1)$) -- ($(2, \y2)$);
   
  \end{tikzpicture}

 \caption{Phases of Cyber Kill Chain} \label{fig:1}
\vspace{-.75cm}
\end{figure}

To analyze such attacks cyber kill chain provides a framework to breakdown the complicated attack into mutually nonexclusive stages or layers. Such a layered approach will enable the analysts to tackle smaller and easier problems at the same time and it will also help the defenders to subvert each phase by developing defenses and mitigation for each of the phases. Cyber Kill chain mainly consists of 7  phases \cite{cyberkillchain:6}\cite{cyberkillchain:7}\cite{cyberkillchain:9}  as shown in Fig. \ref{fig:1}.

There are many articles \cite{cyberkillchain:3}\cite{cyberkillchain:5}\cite{cyberkillchain:8} which describe the cyber kill chain in detail w.r.t. recent attacks but most of them don't discuss the tools and technologies used by the attacker at each stage of the cyber kill chain. In next section we will go through technical aspects involved at each step w.r.t to the attacker's perspective and will see the variety of tools and methodologies utilized by attacker.

\section{Technical Aspects of Cyber Kill Chain}
As described  cyber kill chain defines the flow of a cyber attack and in this 7 layer model each layer is critical. Studying the cyber kill chain will help cyber threats to be identified or mitigated at any layer of attack. Sooner the detection is done lesser is the loss to the organization under attack. This section broadly outlines the technical methodologies, implementation, research, and tools involved at each stage of the cyber kill chain with examples of famous cyber attacks and malwares used in the bygone years.  
\subsection{Reconnaissance}
Reconnaissance means gathering information about the potential target. Target can be an individual or an organizational entity. Reconnaissance can further be broken down to target identification, selection and profiling. Reconnaissance in cyber space mainly includes crawling World Wide Web such as internet websites, conferences, blogs, social relationship, mailing lists and network tracing tools to get information about target. Information gathered from reconnaissance is used in later stages of cyber kill chain to design and deliver the payload. 
Reconnaissance is divided into 2 parts \cite{Reconnaissance:1} (expected to be done in order):

\begin{enumerate}
\item{\textbf{Passive Reconnaissance:}It is gathering the information about target without letting him know about it. }
\item{\textbf{Active Reconnaissance:}It is much deeper profiling of target which might trigger alert to the target. }
\end{enumerate}
\begin{center}
\begin{table}[h!]
\small
\centering

    \begin{tabular}{ |  p{.5cm} |  p{2cm} |  p{1.5cm} | p{3cm} |}
    
    \hline
    & \textbf{Reconnaissance Techniques} & \textbf{Type of Reconnaissance} & \textbf{Techniques Used}\cite{Reconnaissance:2} \cite{Reconnaissance:3}\\ \hline
    1& Target Identification and Selection & passive & Domain Names, whois, records from APNIC, RIPE, ARIN \\ \hline
    2 &Target Profiling  &  &  \\ \hline
     & (a) Target Social Profiling & Passive & Social Networks, Public Documents, Reports and Corporate Websites \\  \cline{2-3}  \cline{3-4} 
     & (b) Target System Profiling & Active & Pingsweeps, Fingerprinting, Port Scanning and services  \\    \hline
    3 & Target Validation & Active & SPAM Messages, Phishing Mails and Social Engineering \\    \hline
  
    \end{tabular}
 \caption{Reconnaissance Techniques}\label{table:1}
\vspace{-1cm}
\end{table}
\end{center}
 Reconnaissance provides knowledge about potential targets which will enable the attacker to decide the type of weapon suitable for target, type of delivery methods possible (Table ~\ref{table:3}), malware installation difficulties and security mechanisms that need to be bypassed. We will now see how the information gathered by reconnaissance is used to develop sophisticated malware.

\subsection{Weaponize}
Weaponize stage of the cyber kill chain deals with designing a backdoor and a penetration plan, utilizing the information gathered from reconnaissance, to enable successful delivery of the backdoor. Technically it is binding software/application exploits with a remote access tool (RAT). Weaponizing involves design and development of the following two components:-\\

\subsubsection{\textbf{RAT (Remote Access Tool)}}
RAT is piece of software which executes on target's system and give remote, hidden and undetected access to the attacker. RAT is usually called the payload of a cyber-weapon. The target system can be a computer, mobile\cite{Weaponize:4} or any embedded device provided the RAT software is properly compiled for the architecture being targeted. The types of access provided by a typical RAT are system explore, File upload or download, remote file execution, keylogs, screen capture, webcam or system power on/off with limited or user level privileges. If by some mechanism the RAT gets administrator/root access then it could include network spreading \cite{Weaponize:2} or network data capture access or persistent installation of anti-detection module. RAT again constitutes of two major parts:
\paragraph{\textbf{Client}}Client is the piece of code which is delivered to the target, executes and creates connection to the Command and Control infrastructure of the RAT. After establishing connection, the client receives the command from its controller. Client in turn executes the command and sends the results back. It is not always necessary to have all RAT functionalities in a single deliverable. There are many instances of RAT like Carberp \cite{Weaponize:7}, Ventir Trojan \cite{Weaponize:8},Poison Ivy \cite{ar11} etc. where RAT functions are delivered to the target in a modular manner by using a very basic stub. The complied code can be deployed in form of binaries or shellcodes. Shellcode in simplistic terms is position independent machine code  generally compiled by assemblers. There also exists techniques \cite{Weaponize:9} which use specially crafted C-language source codes to generate shellcode.
\paragraph{\textbf{Server}} Server is the other half of RAT which runs on the Command and Control infrastructure with nice UI, displays the connection information from target's client part. Server typically has options of commands like keylogger, file browser, screen capture etc. Based on objective attacker gives the command to the client using server interface which is executed by client and the output is returned back to the server. Depending on execution permission level being allotted to the client part these commands can be executed to get full access of target's system. 

While developing RATs major constraints are size, Anti-virus detection, extendibility, and scalability and user friendly interface. Ease of use of a RAT is defined by the server UI, hence a big part of RAT programing is dedicated to UI design of server. 

It is not necessary to code client and server in same language. These days server are using either scripting languages(PHP, Javascript) for web based UI \cite{Weaponize:7} \cite{ar10} or C++,Java, delphi  \cite{Weaponize:10}\cite{ar11} for application based interface. 

\begin{center}
\begin{table}[h!]
\small

\centering
    \begin{tabular}{  |  p{2.0cm} ||  p{3cm} |}
    \hline
    \textbf{ Famous RATs} & \textbf{Famous Exploit Kits} \cite{Weaponize:4} \\ \hline
     Blackshades & Blackhole \\ \hline
    DarkComet  & Nuclear  \\ \hline
     Poison ivy &  Styx \\    \hline
     Bozok & Redkit  \\    \hline
     Njrat & Sweet Orange  \\    \hline
    Apocalypse & Infinity  \\    \hline
    \end{tabular}
  \caption{RATs and Exploit Kits} \label{table:2}
\vspace{-1cm}
\end{table}
\end{center}
\subsubsection{\textbf{Exploit}}
Exploit is the part of weapon which facilitates the RAT to execute. Exploit acts as a carrier for RAT and uses system/software vulnerability to drop and execute RAT. Major objective to use exploit is to evade user attention while establishing a silent backdoor access using RAT. 
Exploits can be of many form like MS Office documents (.doc/ppt) e.g. CVE-2010-3333, CVE-2014-4114, PDF Documents e.g. CVE-2014-9165, CVE-2013-2729, audio/video file e.g. CVE 2013-3245 or web page e.g.CVE-2012-1876, CVE-2014-6332 \cite{Weaponize:11} etc. If target opens any of such files using vulnerable software RAT will be installed on target’s system. After getting RAT installed more exploits like privilege escalation exploits like CVE-2015-002, CVE-2013-3660  are used on target to get higher privileges which can be used for spreading of RAT, persistent access or for destruction of complete system.

Without exploit also there are methods to compromise the target using only RAT but those methods are highly unreliable and not effective these days because of security awareness among users. Embedding RAT in legitimate software executable, sharing via social engineering and faking RAT as genuine image/audio/video files are some of such methods which used in wild in previous years\cite{ar10}.     

\subsection{Delivery}
Delivery is the critical part of cyber kill chain which is responsible for an efficient and effective cyber-attack. For any cyber-attack it is preferable to have target information to ensure a successful attack. In most of the cyber-attacks it is mandatory to have some kind of user interaction like downloading and executing malicious files or visiting malicious web pages on internet. There are some attacks which are performed without user interaction by exploiting network devices e.g. CVE-2014-3306, CVE-2014-9583\cite{Weaponize:11}. When it is clear that target has to interact, it is obvious that attack should have ability to attract and entice the target to interact. This affinity comes from the active and passive reconnaissance.

Delivery is a high risk task for attacker because delivery leaves traces. Therefore most of the attacks are performed anonymously using paid anonymous services, compromised websites and compromised email accounts.

 While delivering the weapon multiple delivery methods are also used because no single method can guarantee 100\% success. Failed attacks sometimes are very useful to get basic information about target's system information.  Such types of information gathering mechanisms are very common in browser based attack where user visits the malicious web page which first try to get user system information and accordingly deliver the weapon\cite{Delivery:1}. 
\begin{center}
\begin{table}[h!]
\small
\centering

    \begin{tabular}{ |  p{0.5cm} |  p{2.5cm} |  p{4.5cm} |}
    \hline
    & Delivery Mechanism  \cite{Weaponize:2}\cite{Weaponize:3} & Characteristics \\ \hline
    1& Email Attachments & Email content is composed to entice the user by using appealing content\\ \hline
     3 & Phishing Attacks & Sensitive information like usernames, passwords, credit card details etc. are extracted by masquerading a trustworthy entity in communication.\cite{Delivery:6} \\    \hline
    4 & Drive by Download & Target is forced to download appealing malicious content from internet. Malicious content could be a image file, pdf/word document or software setup file  \\    \hline
    5 & USB/Removal Media & Infected files are kept in Removable media which afterwards silently infects other systems opening the files.  \\    \hline
    6 & DNS Cache Poisoning & Vulnerabilities in DNS are exploited to divert internet traffic from legitimate servers to attacker controlled destinations.   \\    \hline
    \end{tabular}
\caption{Delivery Mechanisms}\label{table:3}
\vspace{-.8cm}
\end{table}
\end{center}

\subsection{Exploitation}
After delivering the cyber weapon, then as expected the target completes the required user interaction and weapon executes at the target side. On execution, the next step is triggering the exploit. The objective of an exploit is to silently install/execute the payload. To trigger the exploit there are certain conditions that need to be matched:
\begin{enumerate}
\item{User must be using the software/Operating System for which exploit has been created}
\item{The software/Operating System should not be updated or upgraded to the versions wherein exploit doesn’t work.}
\item{Anti-Viruses or any other security mechanism should not detect the exploit or payload neither in statically nor dynamically during run time.}
\end{enumerate}
If all these conditions are fulfilled then exploit is triggered and it will successfully install/execute the payload in target's system. Payload will connect to its Command and Control counterpart to inform about successful execution and wait for further commands to execute.

It is clear that exploit is the most critical part of the chain technically. What is an exploit and how does attacker find such exploits is the next question here.

Exploits are made using vulnerability of softwares publicly identified as CVE\cite{Weaponize:11}. Vulnerability is the software bug which can result in potential threat to the system. And a bug is an unexpected condition in which a program/software misbehaves. Generally while programming lot of efforts is made to avoid such conditions by writing every possible use case, exception handling for all types of inputs, blocking invalid input etc. But most of the times not all cases are covered because of vast variety of user input/interaction. Therefore such invalid or wrong input forces the program to misbehave and this misbehavior is defined as type of vulnerability as mentioned in Table~\ref{table:4}.

Vulnerability are further analyzed by exploit writers to see the possibility to carry and execute a payload. Not all vulnerabilities are exploitable. Some of them are just crashes or DoS or limited execution of program. It means not all vulnerabilities result into exploitable crashes. 
\begin{center}
\begin{table}[h!]
\small
\centering

    \begin{tabular}{ |  p{0.5cm} |  p{1.5cm} |  p{2.5cm} || p{3cm} |}
    \hline
    &Category of Exploits & Type of Exploit & Type of Vulnerability \cite{Delivery:2}  \\ \hline
    1& Operating System Level Exploits & Kernel Exploits, Device Driver Exploits & \begin{itemize} \item{ Denial-of-Services}    \item{ Remote or Local Code Execution} \end{itemize}  \\ \cline{1-2}\cline{2-3}
    2 &Network Level Exploits  & Protocol exploits for FTP, SMTP, NTP, SSH, Router exploits,  &  \begin{itemize} \item{ Privilege Escalation}  \end{itemize}\\  \cline{1-2}\cline{2-3}
    3 & Application/ Software Exploits & Browser Exploits, MS Office exploits, PDF Exploits, Java/Flash Exploits & Memory Corruption: \begin{itemize}\item{Dangling Pointer} \item{Buffer Overflow} \item{Use-After-Free} \end{itemize}\\    \hline
    
    \end{tabular}
\caption{Exploits}\label{table:4}
\vspace{-.7cm}
\end{table}
\end{center}
Regarding the interest that how these vulnerabilities are discovered following papers\cite{Delivery:3} \cite{Delivery:4} \cite{Delivery:5} describe details about fuzzing. Fuzzing is methodology to give customized input to the program and monitoring the output for abnormal behavior. This abnormal behavior is further analyzed w.r.t. given input to describe vulnerability and so to create exploits. 

As mentioned in Delivery sectio sometimes just one delivery method is not sufficient, in the same way not just one exploit is sufficient to attack multiple users specifically during mass attacks. Generally combination of exploits, called an Exploit Kit, is used for this  purpose. As the name suggests Exploit Kit is a collection of multiple exploits for various versions of software. As an example for browser based attacks an exploit kit may have exploits for various versions of Google Chrome, Firefox and internet explorer. During delivery if target uses vulnerable version of any of listed browser for which exploit is available in exploit kit it will be delivered accordingly.

\subsection{Installation}
Host based security measures have grown leaps and bounds compared to other security mechanisms. This in turn has spurned innovation in procedures that circumvent host-based security controls to install, update and regulate the control of the malware installed upon the victims computing device. Traditionally, a computer would become infected by an infection vector like  infected removable media, which in turn will leave a malware executable in some unusual location and modify registry/startup settings so that the malware executable is run everytime the computer boots up. Some user eventually will report this executable to antivirus vendor, who in turn will analyze it and come up with a signature to detect it and in some cases a removal tool. Modern malwares are not that simple anymore \cite{ar16}. Malware nowadays are multi staged and they heavily rely on droppers and downloaders to deliver the malware modules in a much more sophisticated manner. 
\begin{itemize}
\item{\textbf{Dropper} is a program that will install and run the malware to a target system. Before executing the malware code, dropper nowadays tries disabling host based security controls at the target and hides the installed malware.}
\item{\textbf{Downloaders} were designed to perform the same actions as Droppers – disabling the victims’ security and monitoring software, hiding core components and obfuscating the infection vector, etc. – but tended to be smaller than Droppers because they did not contain the core malicious library components. Instead of unpacking an embedded copy of the core malware agent, the downloader would connect to a remote file repository and download the core components.}
\end{itemize}
Today's installation life-cycle incorporates many checks, balances and resilience features – as a means of maximizing the success of the installation, and protecting the participating attackers. Following are some techniques that malware authors use for covert persistent and anonymous installations.
\begin{itemize}
\item{\textbf{Anti-Debugger and Anti-Emulation:} Dropper and downloader components are typically “armored”. Using a variety of packers, crypters and inspection-detection engines, malware authors can ensure that common debugger and emulation analysis techniques will not work. The addition of advanced anti-virtual machine analysis technologies also deters malware analysis to a high extent.}
\item{\textbf{Anti-Antivirus:} Many malware packages include toolsets for automatically disabling host-based detection technologies – disabling anti-virus and IDS products installed on the victim's computer, changing local DNS settings to ensure that no future updates to the operating system or packages are possible, and adding tools that recheck and re-disable protection settings frequently.}
\item{\textbf{Rootkit and Bootkit Installation:} Rootkits are programs which hide the executed payload. Payload file hiding, process hiding are the core functionalities of a rootkit e.g. LRK,AFX,Mebroot \cite{Weaponize:6}. Similarly Bootkits are malwares that are able to hook and patch system to get loaded into the system Kernel, and thus getting unrestricted access to the entire system. Bootkits like Stoned Bootkit \cite{Installation:1} modify the MBR or boot sector for its execution to avoid protections from operating system.   }
\item{\textbf{Targeted Delivery:}By performing a quick inventory of the victim's machine at the dropper/downloader stage and submitting this information to the malware distribution site, the attackers can verify that the compromised computer is real (and not some analysis system) and respond accordingly. In some cases, upon discovering that the “victim” was faked or is an automated analysis system, the attackers would not serve the core malware.}
\item{\textbf{Host-Based Encrypted Data Exfiltration:} Most malware does not encrypt outbound network communication. Critical data stolen from the victim's computer is typically packed and file-encrypted at the host-level before sending over a “clear text” network protocol such as HTTP and SMTP – thereby evading anomaly detection systems and data-leakage prevention systems (DLP)\cite{ar16} }
\end{itemize}
\subsection{Command and Control}
An important part of the remotely executed cyber-attacks is the Command and Control(C\&C) system. C\&C system is used to give remote covert instructions to compromised machines. It also acts as the place where all data can be exfiltrated. Over the years, the architecture of C\&C channels have evolved exponentially owing to the exponential development of defensive mechanisms, namely antiviruses, firewalls, IDSs, etc \cite{ar01}. 
There are mainly three type of C\&C communication structures, namely the traditional centralized structure, the newer peer-to-peer decentralized architecture and the latest Social Networks based structure.
\begin{itemize}
\item{\textbf{Centralized Structure:} Traditionally malware depended upon a classical client server model wherein a central server is used to command and control the infected machines. Since there is only one server, its easy to manage. Also there is no dependence on the infected machines to relay command control signals. Hence failure of random infected machines won’t affect C\&C architecture. But the number of bots that can be controlled depends is constrained by the hardware/software resources available to the C\&C server. Also taking down the server strategically will shutdown the entire C\&C infrastructure. }
\item{\textbf{Decentralized Structure:}Owing to the fact that centralized architecture can be taken down easily and cannot control a large botnet, malware authors started using peer-to-peer P2P architecture for command-and-control. The main aims of using this architecture are scalability (infected machines are used as nodes, and each node in turn is responsible only for a subset of the total botnet), fault tolerance (redundant communication links can be formed to route information) and P2P nature (decentralized architecture removes the significant single point dependence of centralized architectures). Variety of options for P2P e.g. Bit torrent, Gnutella, Kademilia etc. indicates the substantive depth of design possibilities. Eg. Storm \cite{ar02}}
\item{\textbf{Social Networks Based Structure:}Social networks now play a huge part in many people’s lives. Facebook has 1.35 billion registered users as of third quarter of 2014 \cite{ar03}. Most of the social network services are free to use and these services are deemed to be benign in most organizational security policies. Owing to these reasons, social media has now become a viable option for malware authors. These high availability and reliable social networks are used to pass on information in a centralized/decentralized way to the infected machines.  Eg. Taidoor \cite{ar04}}
\end{itemize}

C\&C Communication Traffic analysis is a traditional technique to detect communication pattern among infected machines. As a result malware creators have adopted techniques to hide communication patterns. Anonymous communication techniques involve creating an unobservable communication channel that is resistant to traffic analysis. Unobservable communication channel refers to the communication capability that a third party cannot distinguish between a communication and non-communicating entity. It will be indistinguishable from legitimate traffic. Following is a review of techniques employed by the malware creators to achieve unobservable anonymous communication channel:
\begin{itemize}
\item{\textbf{IRC Chats:}Internet Relay Chat(IRC)\cite{ar05} was developed in 1988 and is a protocol used for text chat over the internet. Its primary function is to provide “channels” which are chat rooms allowing for public/private conversations. These channels were exploited by malwares to send and receive information once infection takes place.}
\item{\textbf{TCP/HTTP/FTP:} Malwares relying on IRC as C\&C channel were now easier to detect at network gateways. Malware authors circumvented this by employing TCP/HTTP/FTP\cite{ar06} \cite{ar07} \cite{ar08} as C\&C channels. Since majority of applications in all the operating systems use these channels for benign communication, malware traffic becomes indistinguishable. Examples : Zeus, Poison Ivy\cite{ar10} \cite{ar11} }
\item{\textbf{Steganography:}Steganography technique involves hiding data inside images, video, audio or any such content which are hosted in attacker controlled harmless websites. The data is then hidden in specially constructed and encrypted inline annotations which are strewn randomly all over the page or extra spurious bytes of images. This hidden data can contain C\&C commands like ip addresses of C\&C server, shellcode commands to execute etc }
\item{\textbf{TOR:}Tor (originally The Onion Router) is a service used to provide anonymity over the internet[ar09]. Tor directs Internet traffic through a free, worldwide, volunteer network consisting of more than six thousand relays to conceal a user's location and usage from anyone conducting network surveillance or traffic analysis.. Apart from providing an anonymity network, Tor also has Hidden Service Protocol which makes it possible for users to hide their location while offering various kinds of public services.. Hidden services work by setting up Rendezvous points which are identified by “.onion” links. SDC Botnet \cite{ar12} is a recent example of malwares employing TOR as C\&C channel.}
\end{itemize}
Attackers would like to stay undetected and covert for as long as possible. Following are 3 techniques commonly used by malware authors to hide their C\&C server from the defense mechanisms of any organization:
\begin{itemize}
\item{\textbf{DNS Fast Flux:}The Domain Name Service(DNS)\cite{ar13}  is a naming system of computers on the internet. DNS is not required to match hostnames to ip addresses in a one-to-one fashion. A single hostname may correspond to many IP addresses to facilitate fault tolerance and load distribution  Malware authors use a technique called fast-flux to hide C\&C locations behind an rapidly changing(fast-flux) network of infected machines. This is achieved by setting a very short TTL(time-to-live) to DNS response which is known as single flux. Another technique called double flux is used which works by switching rapidly through multiple infected nodes’ addresses which in turn act as a proxy relay location to the original C\&C Site.}
\item{\textbf{DNS as a medium:}It is also possible to use DNS system as a communication mechanism rather than using it for setting up the channel. RDATA field of a DNS response can be of variable length and multiple formats. One of these is TXT, which can be used to transfer actual text. The commands are can be encoded into base64 and sent over the RDATA field.  Feederbot\cite{ar15} is one such example.}
\item{\textbf{Domain Generation Algorithms:} Another method to stay under the radar is DGA. DGA programmatically generates pseudo random domain names which is used as address of C\&C server. IT is then upto the attacker to ensure he controls the domains that will be generated. Conficker malware for example, generates 250 domain names by using current UTC date as the seed of pseudo random DGA. The same domains are generated every 3 hours. All the 250 domain names are queried until one of them resolves successfully. \cite{ar14}}
\end{itemize}

\subsection{Act on Objectives}
After getting the communication setup with target system attacker executes the commands. The command used by attacker depends on interest of attack. \cite{cyberkillchain:2}
\begin{enumerate}
\item{\textbf{Mass attack:} Objective of mass attack to get as many targets as possible. In mass attack more than a single system multiple systems together are of interest. Most of such attacks aims for getting bank, email, social media and local system administrator credentials\cite{Objectives:1}. Bigger picture of mass attack is called  BOTNets\cite{Weaponize:1}. BOTNets are mainly used for DDos attacks and virtual coin mining. Virtual coin mining harvest the system processors or GPU to generate virtual currency for the attacker. }
\item{\textbf{Targeted attacks:}  Target attacks are more sophisticated and carried with more caution. Most of such attacks are aimed to get confidential or secret information from target system. Data exfiltration and getting user credentials for online accounts are objective of the attacks. Spreading through the network also becomes the primary goal when target is an organization. }
\end{enumerate}
In both types of attack if attack is intended for destructive purpose it may crash system hard drive or device drivers. Attacker may make CPU to use its maximum capability for long time to damage the processor hardware.

\section{Summary}
In this paper cyber kill has been discussed in detail with its technical aspects. The paper gives technical flow from attacker's perspective which may help security researcher to design prevention mechanisms.This paper explores the recent trends in malware  development by reviewing weaponize and installation techniques. Delivery and Exploitation section gives an insight to the importance of software vulnerabilities, both from an attacker and security researcher's perspective, in the current cyberspace.Study of reconnaissance techniques and Command \& Control infrastructure shows how benign features of network protocols are being misused to achieve nefarious end results.






%

\end{document}